\documentclass[namedreferences]{solarphysics} %

\usepackage[optionalrh,solaromanenum]{spr-sola-addons} % For Solar Physics
\usepackage{graphicx} % For eps figures, newer & more powerfull
\usepackage{color} % For color text: \color command
\usepackage{url} % For breaking URLs easily trough lines
\usepackage{rotating}

\newcommand{\aap}{    {\it Astron. Astrophys.}}

\newcommand{\apj}{    {\it Astrophys. J.}}
\newcommand{\apjl}{   {\it Astrophys. J. Lett.}}

\newcommand{\jgr}{    {\it J. Geophys. Res.}}

\newcommand{\nat}{    {\it Nature}}

\newcommand{\pasj}{   {\it Pub. Astron. Soc. Japan}}

\newcommand{\solphys}{{\it Solar Phys.}}
 
\newcommand{\ssr}{    {\it Space Sci. Rev.}} 

\def\testbx{bx}%
\DeclareRobustCommand{\ion}[2]{%
\relax\ifmmode
\ifx\testbx\f@series
{\mathbf{#1\,\mathsc{#2}}}\else
{\mathrm{#1\,\mathsc{#2}}}\fi
\else\textup{#1\,{\mdseries\textsc{#2}}}%
\fi}

\newcommand{\CIII}{\ion{C}{iii}}
\newcommand{\CIV}{\ion{C}{iv}}

\newcommand{\HeII}{\ion{He}{ii}}
\newcommand{\Halpha}{H$\alpha$}
\newcommand{\OVI}{\ion{O}{vi}}

\newcommand{\FeIX}{\ion{Fe}{ix}}

\newcommand{\NeVIII}{\ion{Ne}{viii}}

\newcommand{\kms}{km~s$^{-1}$}

\newcommand\arcsec{arcsec}

\begin{document}

\begin{article}

\begin{opening}

\title{Quiet Sun Explosive Events: Jets, Splashes, and Eruptions}

%\author{ **}
\author{D.~E.~\surname{Innes}\sep
        L.~\surname{Teriaca}
        }
\institute{Max-Planck Institut f\"{u}r Sonnensystemforschung,
 Max-Planck-Str. 2, 37191 Katlenburg-Lindau\
 email: \url{innes@mps.mpg.de} email: \url{teriaca@mps.mpg.de}}

%\email{innes@mps.mpg.de}
%}
\runningauthor{Innes and Teriaca}
\runningtitle{Structure of Explosive Events}

%\date{Received; accepted }

% AA/2010/14786
%\abstract{}{}{}{}{}
% 5 {} token are mandatory

\begin{abstract}
  % context heading (optional)
Explosive events are broad non-Gaussian wings in the line profiles of small
transition region phenomena. Images from the Solar Dynamics Observatory (SDO)
give a first view of the plasma
 dynamics at the sites of explosive events seen in \OVI\ spectra of a region
 %($18\times300$~arcsec$^2$)
 of quiet Sun, taken with the ultraviolet spectrometer SUMER/SOHO.
 %obtained \OVI\ spectra in sit-and-stare mode for a period of two hours.
 %The spectra were coaligned to high cadence SDO/AIA 304 and 171\AA\ images of the surrounding plasma.
 %The sites of explosive events are investigated in the context of line-of-sight SDO/HMI magnetograms and
 %photospheric flows computed by tracking granules in the SDO/HMI continuum images.
Distinct event bursts were seen either at the junction of supergranular
network cells or near emerging flux. Three are described in the context of
their surrounding transition region (304\AA) and coronal (171\AA) activity.
One showed plasma ejections from an isolated pair of sites, with a time lag
of 50~s between events. At the later site, the EUV images show a hot core
surrounded by a small, expanding ring of chromospheric emission which we
interpret as a `splash'. The second explosive event burst was related to flux
cancellation, inferred from SDO/HMI magnetograms, and a coronal dimming
surrounded by a ring of bright EUV emission with explosive events at
positions where the spectrometer slit crossed the bright ring.
 The third series of events occurred at the base of a slow mini-CME.
All events studied here imply jet-like flows probably triggered by magnetic
reconnection at supergranular junctions. Events come from sites close to the
footpoints of jets seen in AIA images, and possibly from the landing site of
induced high velocity flows. They are not caused by rapid rotation in
spicules.

\end{abstract}


\end{opening}

%_______________________________________________________________

\section{Introduction}
Although transition region explosive events are seen very frequently in
spectra of the quiet and active Sun, images of their sites have been
difficult to obtain until recently. Transition region explosive event is the
name given to small (1-5~arcsec), short-lived (1-5~min) non-Gaussian
broadenings in the wings of spectral lines formed around $10^5$~K. They were
discovered in HRTS spectra of transition region lines by \inlinecite{BB83}
and later described in more detail by \inlinecite{DBB89} and
\inlinecite{Dere94}. Since the launch of SUMER/SOHO explosive events have
been seen regularly in lines with formation temperature $6\times10^4$~K ({\it
e.g.} \CIII) to $7\times10^5$~K ({\it e.g.} \NeVIII) \cite{Wilhelm07}. Both
extended red and/or blue wings are observed. Analysis of the line asymmetry
shows a change with ion, position and evolution \cite{DBB89,WEMW02,Mendoza05}
and often a spatial offset between the red and blue wings. Events often come
in bursts \cite{IBGW97a} with 3-5~min between consecutive events
\cite{NIS04}.

The evolution of line profiles across individual events shows that in some
events the underlying flows are similar to bi-directional jets \cite{IIAW97b}
originating in the transition region and ejected down to the chromosphere and
up towards the corona. Examples have also been found that resemble supersonic
up and down flows along loops \cite{Teriaca04} and, at the limb, flows along
macrospicules \cite{Wilhelm00}. Analysis of the energetics of the flows
suggests that they carry equal energy in both directions \cite{WEMW02}. The
energy input sites do not move during the course of the events but expand and
shrink slowly across the surface with a speed about 25~\kms\ \cite{NIS04}.

Explosive events seem to concentrate along the boundaries of the magnetic
network near sites of cancelling/evolving magnetic flux
\cite{PD91,C2etal98,Muglach08,Aiouaz08} and are thus believed to result from
magnetic reconnection at the Sun. Although the line profiles, energetics and
magnetic field evolution can be explained by energy release at a reconnection
site in the transition region \cite{SSM93,Detal91,IIAW97b,IT99,Retal01}, it
has been argued that the transition region reacts with similar signatures of
explosive energy release to microflaring in the corona \cite{KB00} or
reconnection in the photosphere \cite{TRCF99,RT00}. It is also possible that
explosive events are related to chromospheric jets that cause jet-like
brightenings in transition region and coronal images \cite{Depontieu11}. At
the moment, the debate is still open as to where the primary energy input
site is and can probably only be resolved by high resolution simultaneous
spectroscopic imaging from the chromosphere to the corona.

An alternative to the jet/flow explanation is that the red and blue shifts
are due to rotation rather than up and downflows \cite{Curdt11}. Rotation is
the explanation often given for observations of a red-to-blue change in
Doppler shift during raster observations of spicules seen at the limb
\cite{PM98,Kamio10}; however \inlinecite{Xia05} noticed similar red-to-blue
changes in line profiles obtained during sit-and-stare observations of limb
spicules. These are most simply explained as ejected plasma that reaches a
certain height and then falls back along the same path, casting some doubt on
the rotation interpretation. A more compelling argument against rotation as a
general explanation for explosive events, particularly in disk center
observations, is that nearly all observed off-limb jets/spicules have higher
vertical than lateral velocities \cite{Suematsu08,Scull09,Depontieu12}, and
it is unlikely that events seen at disk center will behave differently and
exhibit faster rotation than outflow. In order to distinguish jet-like from
swirling motion co-temporal images of the chromosphere, corona and spectra of
explosive events like the ones reported here are very useful.

Much has been written about the association of flows, observed as Doppler
shifted emission, with coronal brightenings particularly since the CDS/SOHO
spectrometer generally saw brightenings without significant line broadening
\cite{H97,HLBI99,Hetal03,MD03} and SUMER observing at higher resolution saw
bursts of explosive events associated with brightenings
\cite{Innes01,Brkovic04}. Quiet Sun brightenings at transition region
temperatures can be due to many processes \cite{PHB02} and this makes their
interpretation confusing. For example, brightenings have been associated with
eruption at network junctions where mixed-polarity fields are swept-up by
supergranular flows \cite{H97,Potts07,Innes09} and also with flux emergence
\cite{Doyle04,Subramanian08}. \inlinecite{Innes01} showed that strong
explosive events often had jet-like structure in TRACE \CIV\ images but the
TRACE 171\AA\ quiet Sun emission was weak and difficult to interpret. In
active regions strong explosive events have been seen to coincide with sites
of 171\AA\ brightenings at the footpoints and along \Halpha\ jets
\cite{Madjarska09}.

Now with Atmospheric Imaging Assembly (AIA) \cite{Lemen12} images from the Solar Dynamics Observatory (SDO) we have
much higher quality filter images of the sites of explosive events at both transition region and coronal temperatures,
as well as high cadence line-of-sight magnetic field observations and continuum images that allow us to track the
photospheric flows in which the magnetic flux is entrained.

The question to be addressed here is the structure and magnetic environment
of brightenings associated with transition region explosive events. In the
next section we describe the observations and give an overview of the quiet
Sun region and sites of events. We then describe three explosive event bursts
showing in each case how they relate to brightening in 304 and 171\AA, the
underlying magnetic field and photospheric flows.

\section{Observations}
\begin{figure}
   \centering
  \includegraphics[width=\linewidth]{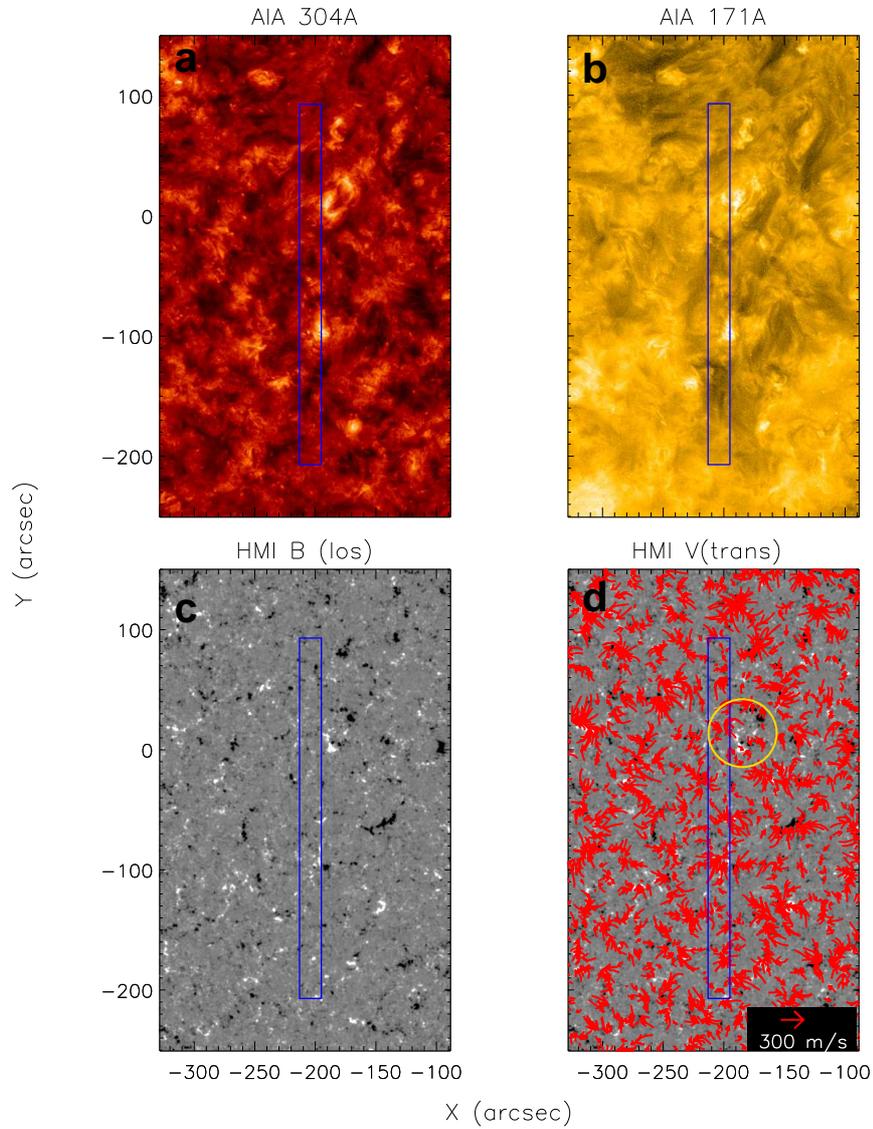}
    \caption{The region observed 2010 June 28 22:11:00~UT: a) 304\AA; b) 171\AA; c) line-of-sight magnetic field, scaled between $\pm 50$~G; d) magnetic field with photospheric transverse flow arrows overplotted. The blue rectangle outlines the region covered by SUMER sit-and-stare observations. The yellow circle in (d) encloses a small site of recent flux emergence. }
\label{over}
\end{figure}

The Solar Ultraviolet Measurements of Emitted Radiation (SUMER) spectrometer
\cite{Wetal95,Wetal97,Letal97} observed the \OVI\ 1032\AA\ line near disk
center in sit-and-stare mode from 22:11~UT 28 June 2010 to 00:11~UT 29 June
2010 through the $4\times300$~arcsec$^2$ slit with a cadence of 60~s. The
Sun's rotation rate at the position of the slit was 0.15~arcsec~min$^{-1}$,
so over the two hours SUMER observed a region $18\times300$~arcsec$^2$.
Chromospheric (304\AA), lower coronal (171\AA), line-of-sight magnetic field,
and supergranular flow images of the observed region are shown in the four
panels of Figure~\ref{over}. The explosive events are discussed in the
context of these four observables obtained from the AIA and Helioseismic and
Magnetic Imager (HMI) \cite{Scherrer12} on SDO.

\begin{figure}
   \centering
  \includegraphics[width=0.8\linewidth]{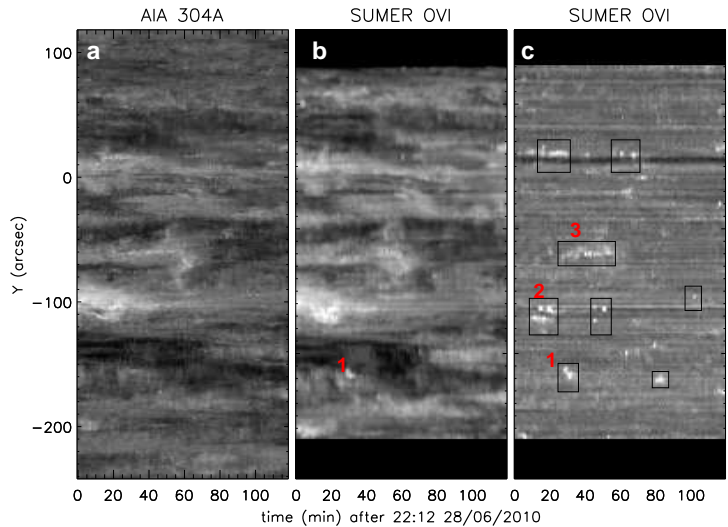}
    \caption{Coaligned AIA 304\AA\ and SUMER \OVI\ time series: (a) 304\AA\ intensity; (b) \OVI\ 1032\AA\ intensity;
    (c) \OVI\ 1032\AA\ Doppler width scaled below 80~\kms. In (c), events bursts are enclosed with black rectangles.
    The ones discussed in the text are numbered. The horizontal stripes in (c) are artifacts due to imperfect flatfielding,
    with the exception of the larger one around solar Y=10, that is due to an MCP defect.}
\label{coalign}
\end{figure}

During our observations the \OVI\ line was dispersed on the bare part of
detector B. After each 60~s exposure 100 pixels across the \OVI\ line were
transmitted to the ground. This corresponds to a spectral window of 4.4\AA\
or $\pm600$~\kms\ in line Doppler shift. After decompression, the SUMER data
have been corrected for dead time losses, local gain depression, flat-field
and geometric distortion using the standard software provided in SolarSoft.
The best available flatfield was obtained almost one year after the time of
these observations. Since the flatfield evolves with time and position on the
detector, we are unable to remove all flatfield artifacts. Residuals from the
flatfield subtraction lead to stripes in maps of fitted parameters ({\it
e.g.}, line width, see Figure~\ref{coalign}). We checked their effect on the
line profiles and found it to be modest. The SUMER detectors also have some
small physical defects (damaged MCP) that cannot be corrected. Spectra from
these locations are unreliable and not considered. Finally, it is important
to mention that count rates in excess of 50 counts~s$^{-1}$ can produce
irreversible gain depression, and thus alter the spectral profiles. This
situation arises during the second event studied, where the line center count
rates are high. However, careful inspection of the raw data reveals that the
explosive event signature in the wings can be trusted.

AIA takes full Sun images with a spatial resolution of 0.6 arcsec pixel$^{-1}$ and time cadence of 12~s in 10 wavelength bands,
including seven extreme ultraviolet (EUV), two ultraviolet and one
visible filter.
For the SDO analysis, we select two channels: 304\AA\ which is dominated by the \HeII\ lines formed around $5\times10^{4}$~K and 171\AA\
which is centered on the \FeIX\ line formed around at $6\times10^{5}$~K.
The level 1 data have been processed to deconvolve the point spread function and then registered to level 1.6 using SolarSoft routines.

The underlying magnetic field and its evolution is obtained from HMI
line-of-sight magnetograms with a pixel size 0.5~arcsec. Supergranule lanes
and junctions are obtained by tracking the horizontal photospheric flows in
HMI continuum images, taken with a cadence of 45~s. Flow velocities have been
computed with the balltracking method developed by \inlinecite{Potts04} (see
also \inlinecite{Attie09}). First the continuum images are de-rotated to the
same time, then tracking is done on individual granules in consecutive
images. Supergranule flows are computed by smoothing the resultant velocities
over a Gaussian width of 4~arcsec, and averaging over 45~min. The positions
of the flow arrows shown in the figures have been computed by integrating
arrow positions along streamlines. New arrows are continually added at random
positions where there are few arrows and removed from regions where they
cluster together. Arrows collect where the flows converge along supergranular
cell boundaries and at junctions. Thus the arrows in Figure~\ref{over}d
mostly overlie the magnetic field concentrations, except where flux emergence
has occurred (yellow circle).

The contribution function of the \OVI\ 1032 line peaks at about
$3\times10^5$~K and has a tail extending to higher temperatures. For this
reason the \OVI\ matches both the 304\AA\ and 171\AA\ structures fairly well.
Although the \OVI\ formation temperature is closest to that of \FeIX, the
transition region dynamics are better reflected in the 304\AA\ images and
these are therefore used for coalignment with SUMER.
%The spatial and
%temporal resolution, and exposure time %differences make an exact one-to-one
%comparison difficult.
In Figure~\ref{coalign}, we have created a synthetic
sit-and-stare 304\AA\ image of the region covered by the SUMER slit. For each
SUMER exposure we have taken the 304\AA\ image closest to the central time
and then binned the data over 4~arcsec in the X direction and 1~arcsec in the
Y direction to give the same spatial coverage as the SUMER slit. There is
very good overall agreement, and the alignment is within one arcsec in both
directions. The main discrepancies are where there are small brightenings in
the \OVI\ (e.g the brightening labelled `1'). As shown in the next section,
these probably arise because the main brightening is very quick and not
present in the 304\AA\ image taken to represent the 1~min SUMER exposure. For
coalignment, we find it sufficient to use one AIA image per minute.

\begin{figure}
   \centering
  \includegraphics[width=0.6\linewidth]{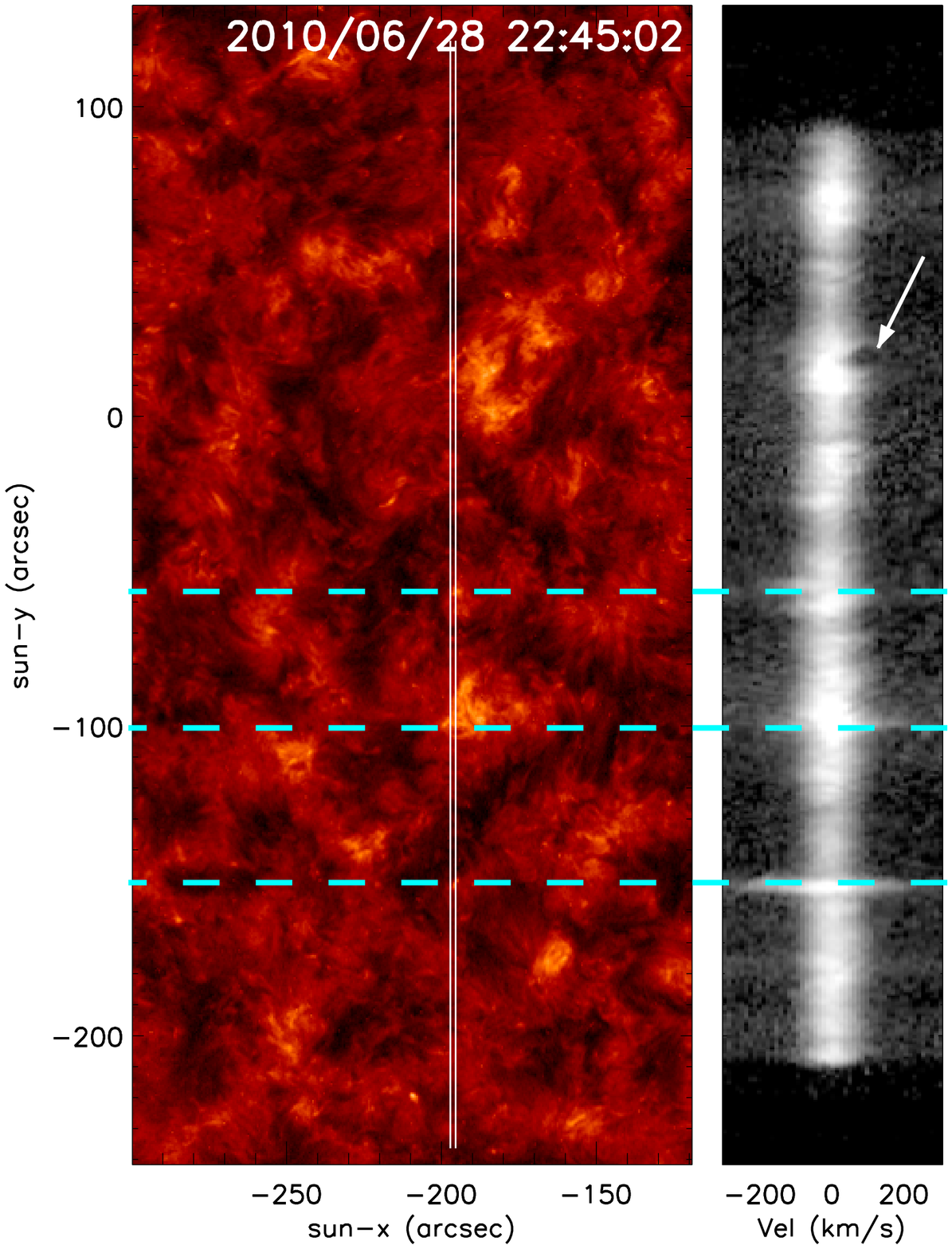}
    \caption{AIA 304\AA\ structure around SUMER \OVI\ explosive events: left) 304\AA\ intensity; right) \OVI\ 1032\AA\ stigmatic spectrum. All three explosive events align with 304\AA\ brightenings (dashed lines). The white arrow points to a detector defect. This is one frame from the accompanying movie AIA\_SUMER.}
\label{movie}
\end{figure}

An example of coaligned SUMER stigmatic spectra with the cotemporal AIA 304\AA\ image is shown in Figure~\ref{movie}.
As indicated by the horizontal dashed lines, explosive events coincide with brightening in the 304\AA\ image.
 This was true for the whole time series, shown in the accompanying movie AIA\_SUMER.
%The full time-series of \OVI\ spectra alongside the AIA 304\AA\ images can
%be seen %Unfortunately a detector defect near the top of the slit distorted
%spectra from the bright region near %$Y=0$~arcsec.
It is also apparent from the movie that explosive events occur at specific
sites and come in bursts. Strong events appear as profiles with widths
greater than 80~\kms. In Figure~\ref{coalign}c, we show the \OVI\ Doppler
widths and enclose explosive events bursts in black rectangles. The grouping
has been made by inspection and based on the temporal and spatial proximity
of events. A single event is one or more consecutive event profiles at a
single site. If several neighbouring sites produce an explosive event, or if
explosive events come and go at a single site then we call this an explosive
event burst.

\section{Event evolutions}
All explosive events seen in \OVI\ were associated with small-scale
brightenings in the AIA 304\AA\ images. Most events occurred in bursts but
there were also several single or paired events. There were four main bursts.
Unfortunately the two northern ones were near the detector defect and so are
excluded from further study. We note that these bursts were just east of the
newly emerged flux (Figure~\ref{over}d). In the following section we show the
environment around the other two bursts and the brightest of the event pairs
to give a feel for the type of structure producing explosive events.

\subsection{Jets and splash at loop footpoints}
\begin{sidewaysfigure}
   \centering
  \includegraphics[width=\textheight]{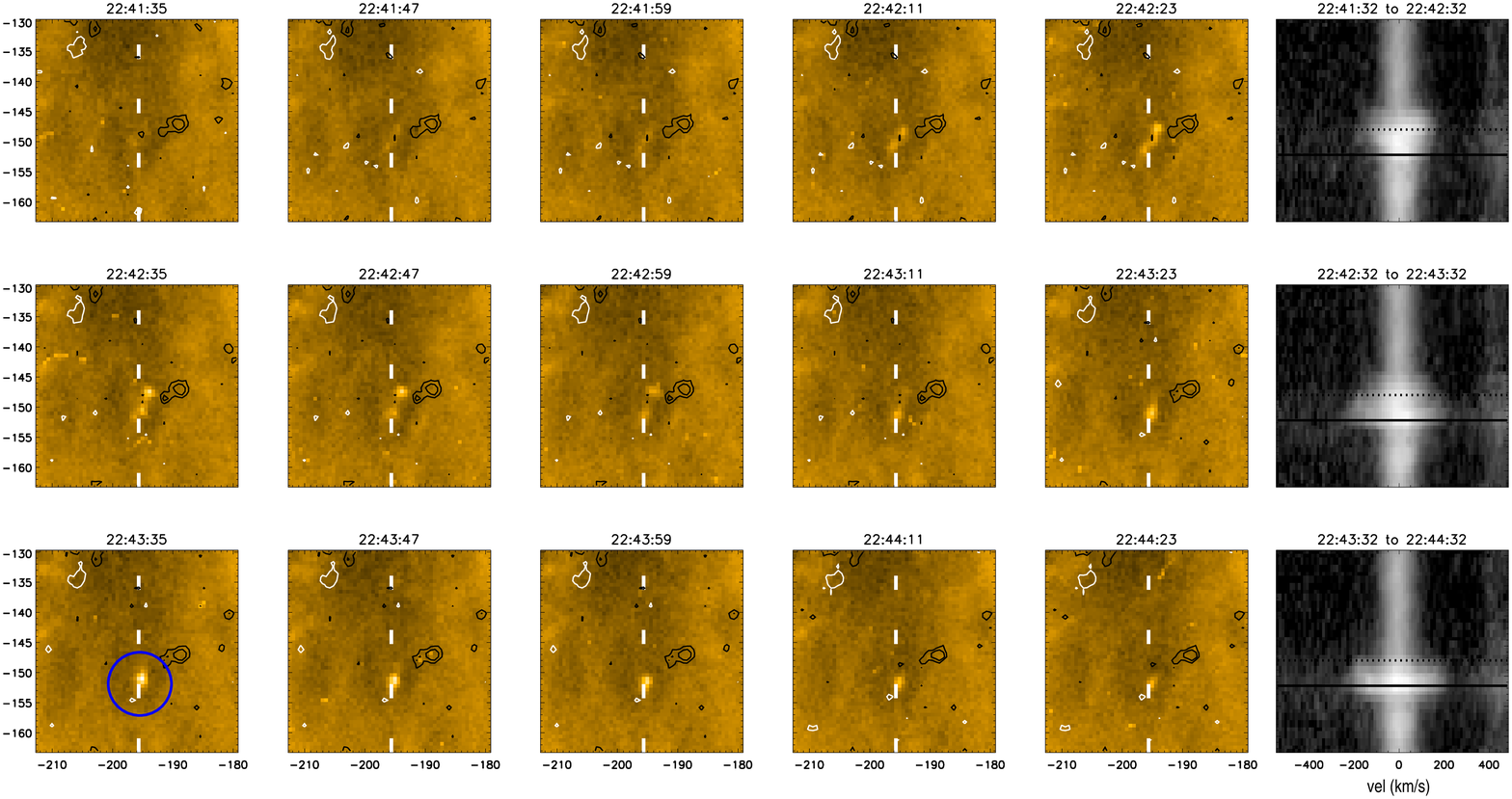}
    \caption{The five 171\AA\ AIA images along each row shows the evolution
    during the SUMER \OVI\ exposure on the right.
    The vertical light blue dashed line in the middle of each 171\AA\ image
 indicates the position of the SUMER slit. Line-of-sight magnetic field
contours at +/-20, 50~G are overplotted in white/black. On the SUMER spectra
horizontal black solid and dotted lines indicate the position of the profiles
shown in Figure~\ref{E1_304}. The blue circle in the bottom row surrounds the
`splash'.} \label{E1_171}
\end{sidewaysfigure}

\begin{sidewaysfigure}
   \centering
  \includegraphics[width=\textheight]{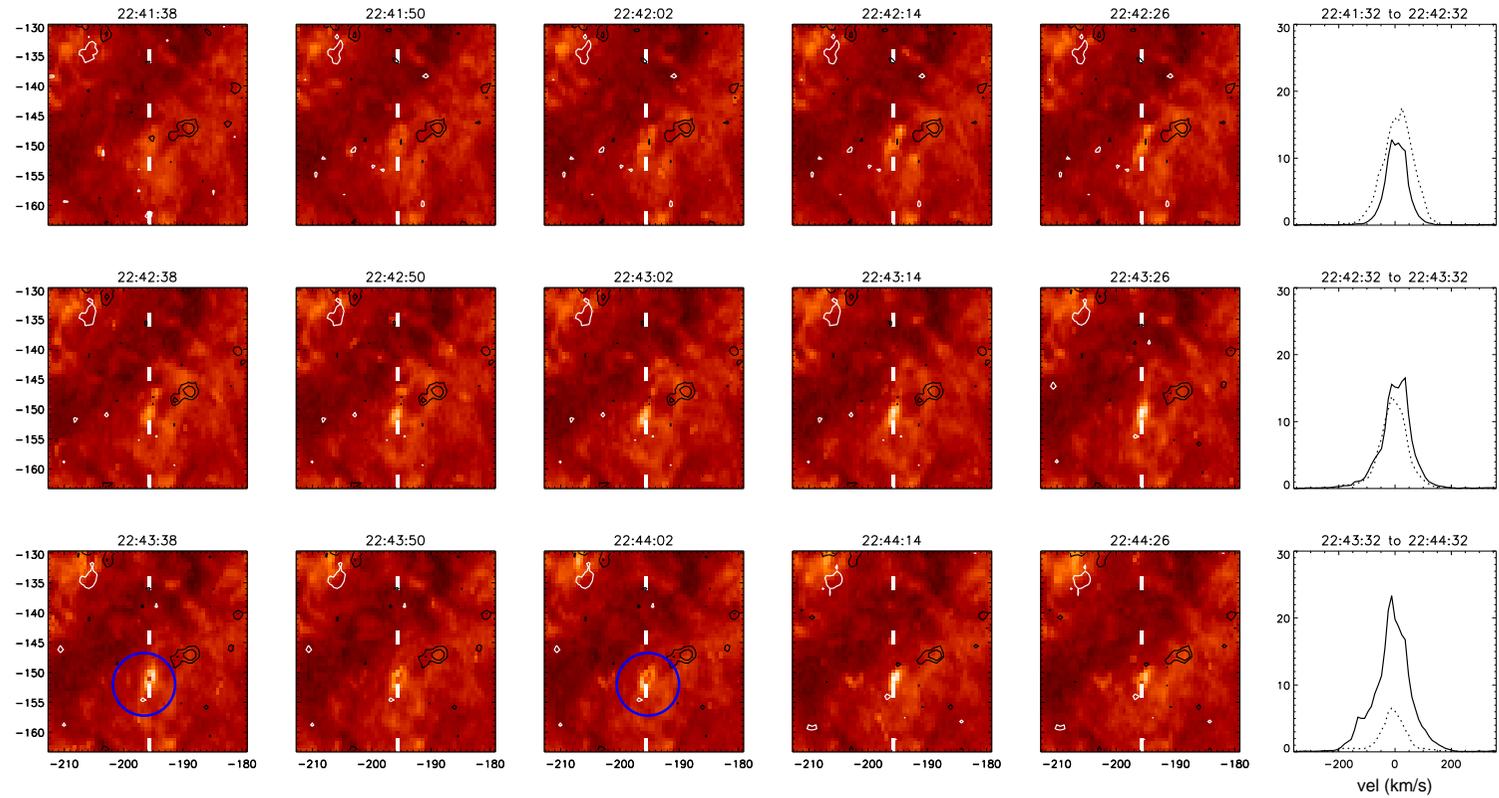}
    \caption{Corresponding AIA images at 304\AA\ of explosive events shown in Figure~\ref{E1_171}.
The spectra on the right have been replaced by line profiles across the
events. The dashed line profile is from the upper event and the solid line is
from the lower one. Blue circles on the bottom row surround the `splash'.}
\label{E1_304}
\end{sidewaysfigure}

This isolated pair of explosive events (`1' in Figure~\ref{coalign}) was
associated with a pair of EUV brightenings that occurred in an otherwise
quiet region, and allows a clear view of the individual explosive event
sites. The evolution shown in Figures~\ref{E1_171} and \ref{E1_304} only
lasted three min. Each row shows the five AIA images from the 171\AA\
(Figure~\ref{E1_171}) or 304\AA\ (Figure~\ref{E1_304}) channel taken during
the 60~s SUMER exposure, shown on the right. The 304\AA\ images are taken 3~s
after the 171\AA\ ones. SUMER stigmatic images are shown on the right of the
171\AA\ images and the line profiles at the sites of brightening on the right
of the 304\AA\ images.

In the AIA images, there are basically two small brightenings. They are seen
as a well-separated pair in the first three frames along the middle row. The
event started around 22:42:11~UT (the fourth frame on the top row) when both
sites appear.
 The northern one is clearly the brightest and remained so for the first $\sim50$~s.
Then it faded and has disappeared in the 22:43:11~UT frame.
 %The brightening at the northern site lasted about60~s.
As the northern site faded, the intensity at the southern site started to
increase and reached its peak brightness at 22:43:35~UT (first frame on the
bottom row). We note that both sites produced explosive events in sync with
the AIA brightenings. Initially only the northern site produced an explosive
event (top row), then both (middle row) and at the end an explosive event was
at the southern site only (bottom row).

The southern site has a distinctly different structure in the 171\AA\ and 304\AA\
 images. The 304\AA\ channel shows a small ring of bright emission with a dark center, while the 171\AA\ channel shows
a small bright point that coincides with the dark center of the 304\AA\ ring (blue circles, Figures~\ref{E1_171} and
\ref{E1_304}). Initially the 304\AA\ ring is only 1-2 pixels (800~km) wide but over the following two images (30~s),
the diameter doubles (first three frames in the bottom row, Figure~\ref{E1_304}).
 %Then at 22:44:14, the ring disappeared and the 304\AA\ brightening co-incided with the fading 171\AA\ %emission.

\begin{figure}
   \centering
  \includegraphics[width=\textwidth]{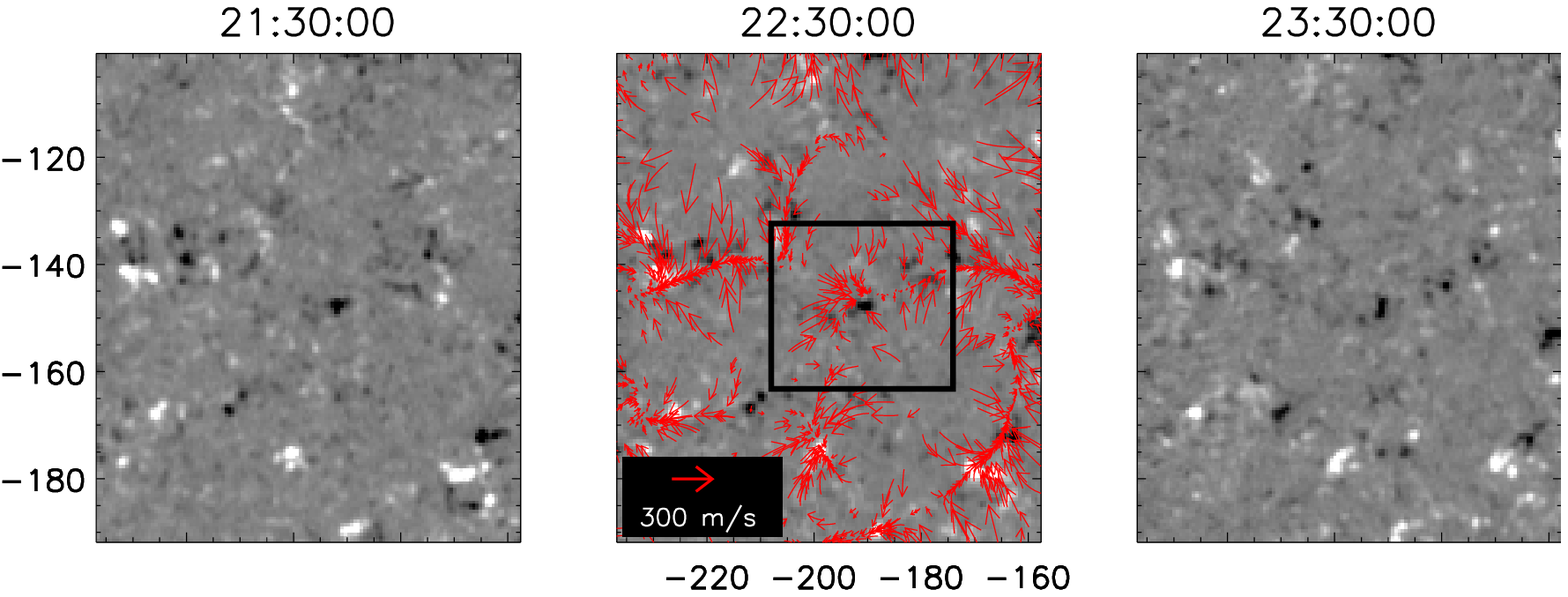}
    \caption{SDO/HMI line-of-sight magnetograms for one hour before, at the time of, and one hour after event `1'.
    The supergranular flows are overplotted as red arrows on the central image.
    The black rectangle surrounds the explosive event site. The magnetic field is scaled $\pm50$~G.}
\label{E1_hmi}
\end{figure}

 The magnetic field evolution and supergranular flows, depicted as arrows, are shown in Figure~\ref{E1_hmi}.
This was a small, almost unipolar (negative) supergranular junction. There
was very little magnetic field evolution for the 2 hours around the event.
HMI magnetograms have a $1\sigma$\ noise level of 10~G \cite{Liu12} so there
may be weak positive polarity fields present.

The structure of the event suggests that this may have been a small magnetic
loop. In particular because the 171\AA\ emission connects the two
brightenings in the frame at 22:42:23~UT (Figure~\ref{E1_171} right hand
image on the top row). Assuming a semi-circular loop, the length would have
been about 5~Mm.
%(4arcsec * pi *800/2)
 At onset (22:42:11~UT) both sites appeared in the same 171 or 304\AA\ image.
We checked the EUV images from other channels and found that the two sites
always appeared in the same image. Initially only the northern one produced
jet-like flows.
 The travel time for a plasma jet with the observed velocity, $\sim100$~\kms,
from one footpoint to the other would have been 50~s which is about the time
between the fading of the northern footpoint and the peak intensity of the
southern one. There was also a delay between the northern and southern
explosive events. Due to the 60~s exposure time of SUMER, it is not possible
to determine how long it was but it seems reasonable to assume it followed
the brightening of the two footpoints.

 A jet-like flow along a loop may explain the hot core surrounded by a ring of
cooler emission seen at the other footpoint because high velocity plasma may
create a splash when hitting the surface. It is not clear how much energy
will be dissipated at the landing site. 1-D numerical simulations of plasmoid
ejections along loops show that reflection can occur at the landing site
\cite{Sterling91}. In 3-D, the situation is likely to be similar but the
extra dimensions may give rise to splashes. Reverse jets have also been seen
on a larger scale in X-ray images \cite{Strong92,Shimojo07}.

 The bright core
was also seen in 211 and 193\AA\ images therefore it probably reached
temperatures at least as high as 2~MK. Hot plasma is sometimes seen at the
base of surges and jets \cite{Madjarska09}, so it is also possible that the
ring of chromospheric emission is the response of the surrounding plasma to a
rapid surge-like up- and downflow.

%\cite{Lee12} between active regions.
%Assuming all energy in the ejected plasmoid is dissipated as heat, then the temperature at the landing site %depends on the ejected plasmoid velocity and the ratio of the density at the footpoint to the plasmoid %density:
%$$
%m_A n_p V_p = n_f k_B T_f
%$$
%\noindent
%where $m_A=2\times10^{-24}$ is the mean atomic mass, $n_p, V_p$ are the plasmoid number density, velocity, %$n_f, T_f$ are the plasmoid number density, temperature, and $k_B$ is the Boltzmann constant. So assuming %$V_p$ is 200~\kms, $T_f \approx 5(n_p/n_f) 10^6$~K.

\subsection{Flux cancellation at supergranular junction}
\begin{sidewaysfigure}
   \centering
  \includegraphics[width=\textheight]{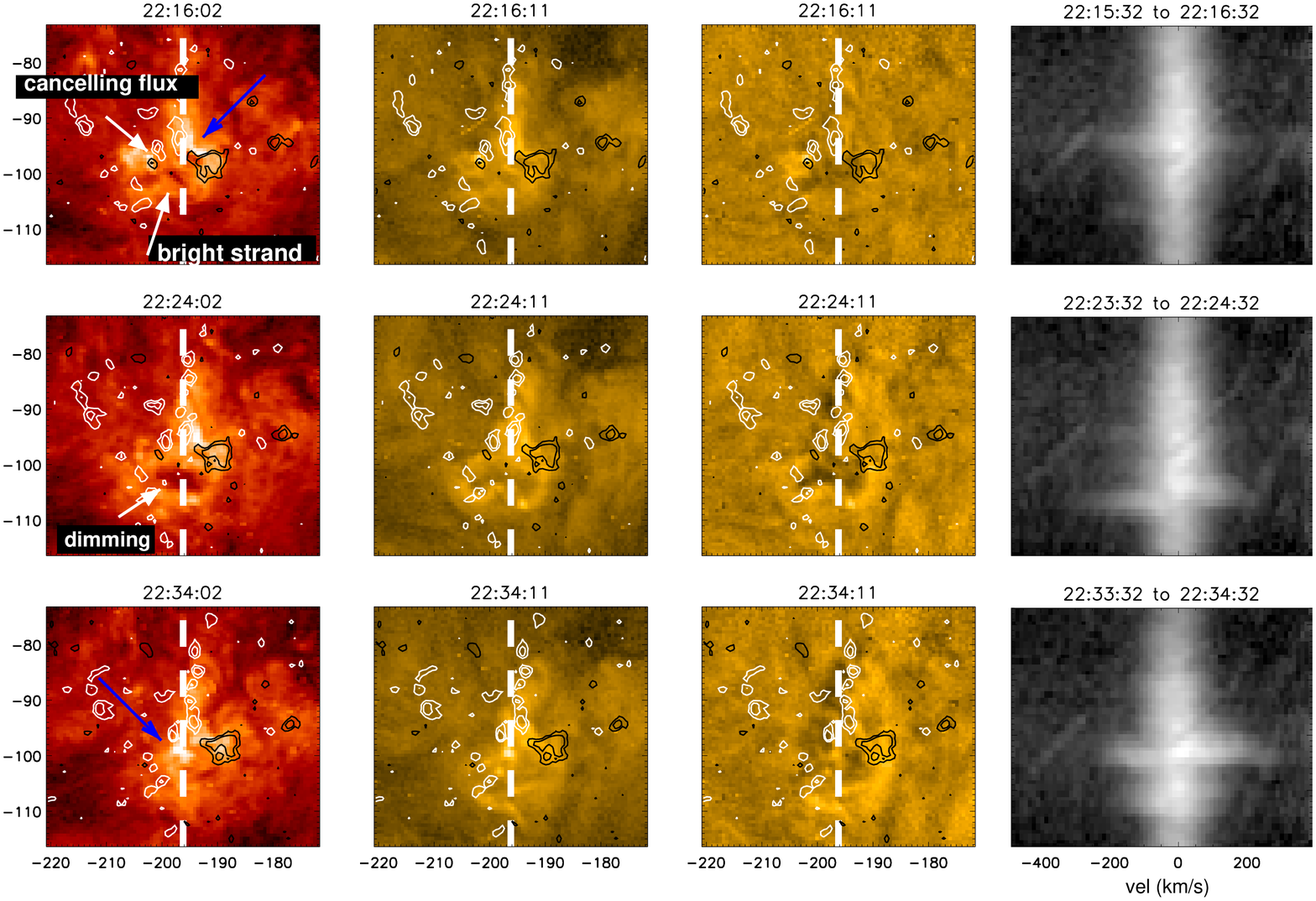}
    \caption{Structure of sites around events `2':(left-to-right) SDO/AIA 304, 171,
    and base difference 171\AA\ images taken mid-way through the 60~s SUMER exposure
    on the right. The base difference is intensity difference between the 171\AA\
    image and an image taken 5~min before the event (i.e. at 22:10~UT).
    Magnetic contours at +/-20, 50~G are overplotted in white/black.
    Blue arrows point to bright emission sites of explosive events.}
\label{E2_aia}
\end{sidewaysfigure}

This burst of explosive events, labelled `2' in Figure~\ref{coalign},
extended over 20~\arcsec, and was seen as the the bright patch of network
near (-200,-100) in Figure~\ref{over} drifted westward under the SUMER slit.
%The evolution
%and association of explosive events to brightenings is best seen in the
%movie AIA\_SUMER.
At this site, there were multiple 304\AA\ brightenings and
dimmings (see the movie AIA\_SUMER). Especially at the time that the region
was scanned by the SUMER slit, there were several sudden 304\AA\ dimmings
with a bright ring around their edge.
% that gave the impression of a small loop-system opening up.
The biggest such dimming was associated with this burst of explosive events,
and is illustrated by snapshots of 304\AA, 171\AA\ and 171\AA\ base
difference images in Figure~\ref{E2_aia}. Unlike the structure described in
`1', the central dimming and outer ring were seen simultaneously at 171\AA,
as well as at 193 and 211\AA\ (not shown). We note that explosive event
sites, indicated with arrows, are where the slit overlaps with the bright
ring.

The ring-like structure suggests eruption of a small loop system with
connections in all directions. Such an eruption would explain the explosive
events along the ring's edge since this is where one expects the footpoints
of the erupting loops. Due to the 60~s time cadence, and sit-and-stare
observing sequence, the explosive event structure cannot be resolved. They
could be caused by bi-directional jets \cite{IIAW97b} or flows along small
loops. The important point here is that they coincide with brightenings along
the edge of the erupting ring so are flows not rotating spicules.

\begin{figure}
   \centering
  \includegraphics[width=\textwidth]{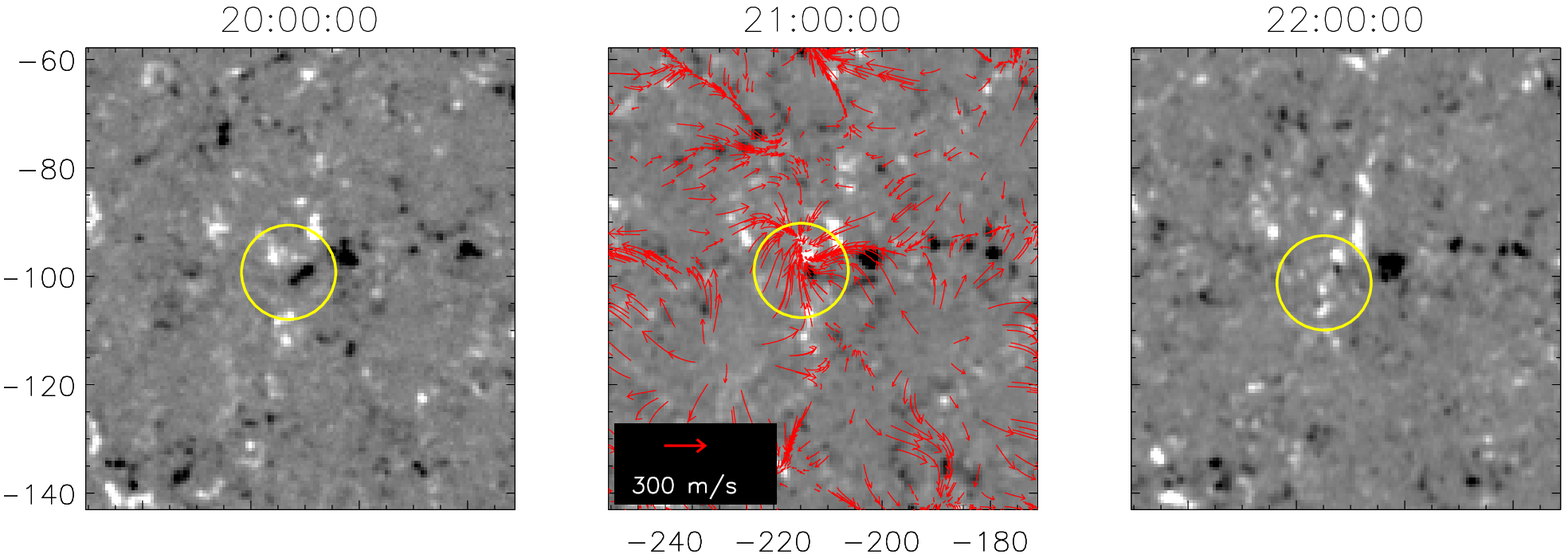}
    \caption{SDO/HMI line-of-sight magnetograms leading up to the burst of events `2'.
    The supergranular flows are overplotted as red arrows on the central image.
    The yellow circle surrounds cancelling flux. The magnetic field is scaled $\pm50$~G.}
\label{E2_hmi}
\end{figure}

As seen in Figure~\ref{E2_hmi} and the movie EVENT2\_HMI, the event occurred
at a mixed-polarity junction where converging flows were driving flux
cancellation (yellow circles). Above we noted that the evolution looked like
the opening of a small loop system and in Figure~\ref{E2_aia} one sees that
initially there was a bright 304\AA\ strand running southwards from the
cancelling flux region. Eruption of this strand seems to have been the cause
of the larger dimming. The actual dynamics of the eruption was probably more
complicated than reconnection at a loop footpoint leading to eruption because
there was a lot of surrounding small-scale activity but it may describe the
essential process. Thus this event is a good example of flux cancellation
leading to eruption and plasma jets in the transition region at the
footpoints of an erupting loop system.


\subsection{Explosive events at the base of a mini-CME}
\begin{sidewaysfigure}
   \centering
  \includegraphics[width=\textheight]{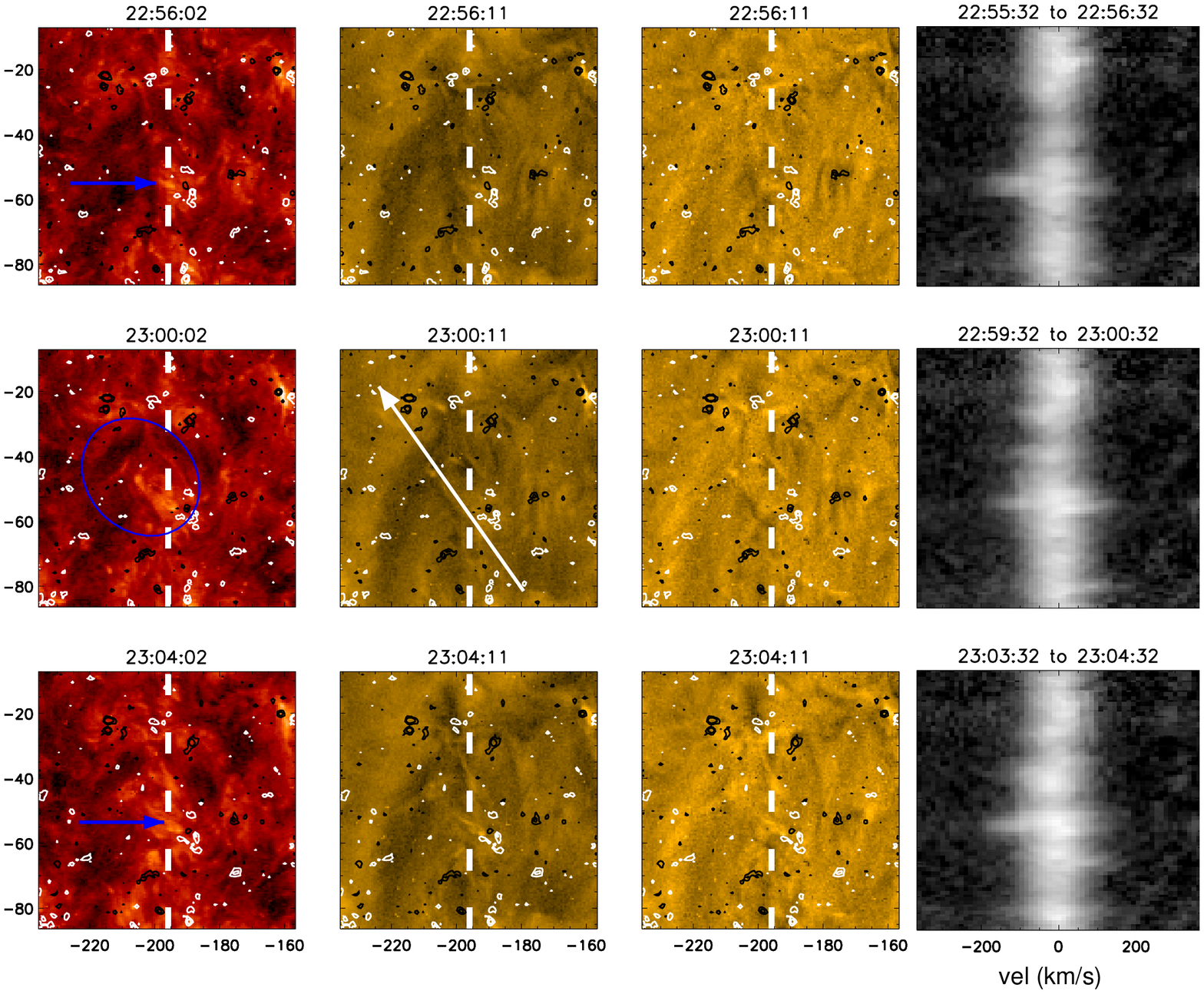}
    \caption{Structure of sites around events `3':(left-to-right) SDO/AIA 304, 171,
    and base difference 171\AA\ images taken mid-way through the 60~s SUMER exposure
    on the right. The base difference is intensity difference between the 171\AA\
    image and an image taken 5~min before the event (i.e. at 22:50~UT).
    Magnetic contours at +/-20, 50~G are overplotted in white/black.
    The white arrow shows the direction and position of the time evolution shown in Figure~\ref{E3_vel}.
    Blue arrows point to the sites of explosive events. The blue circle surrounds the mini-CME eruption.}
\label{E3_aia}
\end{sidewaysfigure}

\begin{figure}
   \centering
  \includegraphics[width=\textwidth]{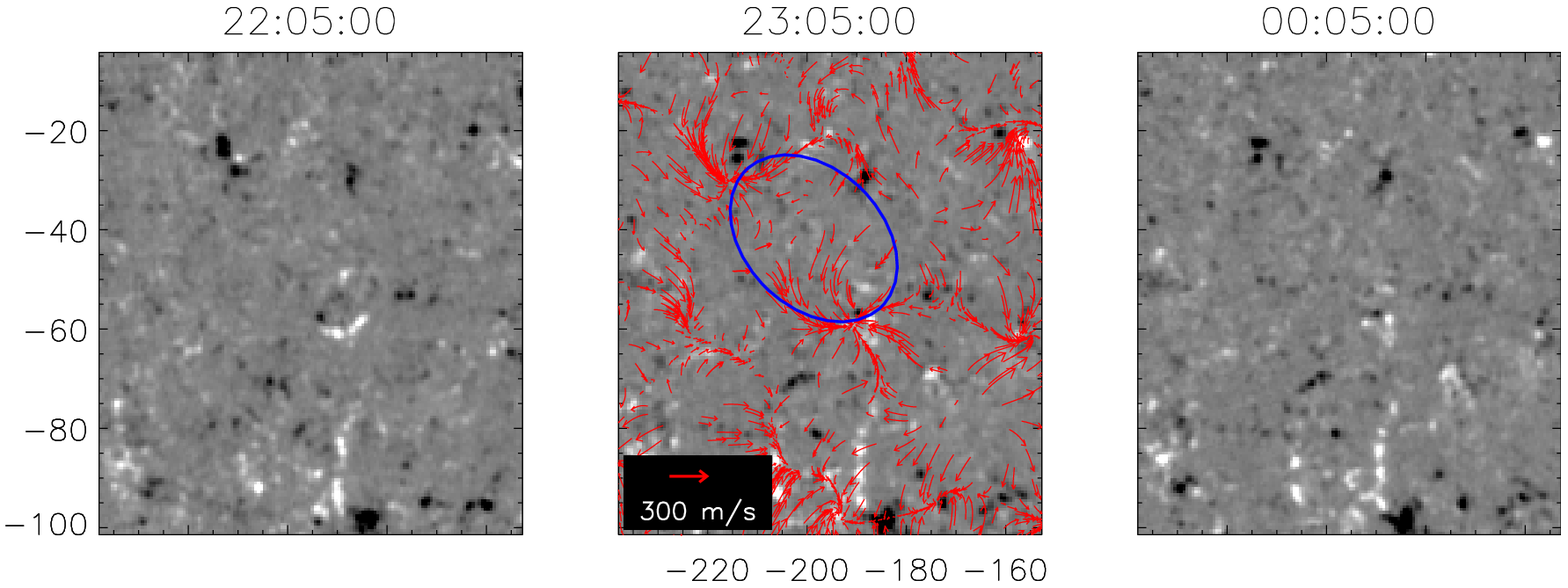}
    \caption{SDO/HMI line-of-sight magnetograms for one hour before, at the time of, and one hour after event `3'.
    The supergranular flows are overplotted as red arrows on the central image. The blue circle indicates the region affected by the eruption.}
\label{E3_hmi}
\end{figure}

This series of explosive events, labelled `3' in Figure~\ref{coalign},
occurred at the onset site of a quiet Sun mini-CME. The eruption is best seen
in the accompanying movies EVENT3\_304 and EVENT3\_171. It started at a small
mixed polarity supergranular junction and propagated across the supergranular
cell to the north east (see blue circles in Figures~\ref{E3_aia} and
\ref{E3_hmi}). The eruption is small compared with examples of mini-CMEs in
the literature \cite{Innes09,Podlad10,Zheng11}. It was the largest crossed by
the slit in the two hours of observation. Significant explosive events are
only seen at the base of the mini-CME where the 304\AA\ brightens, not where
the propagating front moves along the slit.

\begin{figure}
   \centering
  \includegraphics[width=0.6\textwidth]{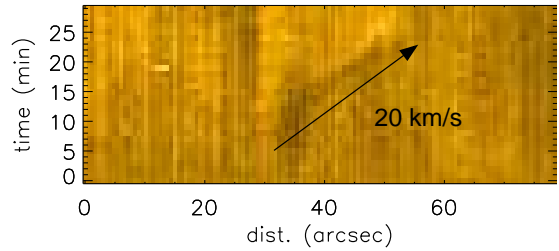}
    \caption{171\AA\ base difference evolution along white arrow in Figure~\ref{E3_aia}.}
\label{E3_vel}
\end{figure}

To determine the speed of the front, we constructed a time series of 171\AA\ base difference images along the white
arrow in Figure~\ref{E3_aia}. The front speed across the supergranule cell was 20~\kms\ (Figure~\ref{E3_vel}). Waves
with velocity 20-40~\kms, emanating from explosive event sites and crossing neighbouring supergranular cells were
picked up by previous TRACE and SUMER co-observations \cite{Innes01}. In that study the waves showed up in SUMER
density sensitive line ratios not in the filter images so they may not be exactly the same phenomena as the front
observed here but the signs are that explosive events commonly give rise to larger-scale waves and propagating fronts.
Therefore we believe explosive events are signatures of plasma jets at the base of eruptions, not the moving front
which explains why they do not move along the slit.

\section{\textbf{Summary}}
This is the first analysis of simultaneous explosive event and SDO imaging
observations. Each event or burst of events have their own unique
characteristics but all explosive events were associated with brightening in
304\AA\ images. They appear to be from quasi-stationary jet-like flows and
are not due to the rapid rotation of spicules. Bursts of explosive events are
most commonly found at mixed-polarity junctions of supergranular cells. A
small region of flux emergence was also the site of multiple explosive events
but this region could not be studied due to a defect on the SUMER detector.
We have investigated one pair of explosive events and two explosive event
bursts.

The pair of explosive events seem to have been produced by flow along a loop.
Both explosive event sites appeared simultaneously in EUV channels but the
time lag of 60~s between their peak brightness and the delay between
explosive events suggests that energy released at one footpoint drove a flow
along the loop to the other footpoint where there was a second explosive
event. The structure of the 304 and 171\AA\ emission at the site of the
second event revealed a hot core surrounded by a ring of chromospheric
emission. Such a structure at the site of an explosive event has so far not
been reported. We interpret it as a splash in which the explosive event is
the downward and reverse jet. Alternatively, it could be the response of the
lower atmosphere to a small-scale spicule-like ejection that later falls back
to the surface.

Both explosive event bursts investigated here were associated with coronal
dimming, implying that the coronal parts of the loops erupted. It is very
likely that explosive events are bi-directional plasma jets at the footpoint
of loops and these are tied to the photosphere explaining why explosive event
sites do not move across the surface. This is consistent with the
co-ordinated \Halpha, 171\AA, and SUMER observations of a surge
\cite{Madjarska09} which showed strong stationary explosive events
co-incident with 171\AA\ brightening at the footpoints of \Halpha\ jets.

We have seen that eruptions associated with explosive events can be
significantly larger than the explosive events themselves. Similar flows
around explosive event sites were noted by \inlinecite{Teriaca04}. When
deducing the energy of the events the full extent of eruptions must be
considered. This will be the topic of future studies.

\begin{acknowledgements} The authors thank the referee for critical and helpful
comments. We would also like to thank R. Attie and A. Genetelli for
discussion on the balltracking method. We are indebted to the SDO/AIA teams
and the German Data Center at MPS for providing the data. The SUMER project
is financially supported by DARA, CNES, NASA and the ESA PRODEX programme
(Swiss contribution).

\end{acknowledgements}
\newpage
\bibliographystyle{spr-mp-sola}

%

%
%\clearpage\Online\appendix
%\section{Additional figures}

\end{article}
\end{document}